\documentclass[]{aa}
\usepackage{graphics}
\begin{document}
\thesaurus{08(08.19.4; 08.19.5 SN 1997D; 09.19.2)}
\title{
  The nature of SN~1997D: low-mass progenitor and weak explosion
  \thanks{All the SN~1997D spectra used in this paper
        were kindly provided by Massimo Turrato}
}
\author{Nikolai N.\,Chugai\inst{1,2}
  \and Victor P.\,Utrobin\inst{3,4}}
\offprints{N.N.\,Chugai (nchugai@inasan.rssi.ru)}
\institute{
  Institute of Astronomy RAS,
  Pyatnitskaya 48, 109017 Moscow, Russia
\and
  European Southern Observatory,
  Karl-Schwarzschild-Str. 2, D-85748 Garching, Germany
\and
  Institute of Theoretical and Experimental Physics,
  B.~Cheremushkinskaya St. 25, 117259 Moscow, Russia
\and
  Max-Planck-Institut f\"ur Astrophysik,
  Karl-Schwarzschild-Str. 1, D-85740 Garching, Germany}
\date{Received / Accepted}
\titlerunning{The nature of SN~1997D}
\authorrunning{N.N.~Chugai \& V.P.~Utrobin}
\maketitle
\begin{abstract}
We analyzed the spectra and light curve of the 
  peculiar type II-P supernova 1997D
  to recover ejecta parameters. 
The optimal hydrodynamical model of SN~1997D, which meets observational 
  constraints at the photospheric epoch, suggests a low explosion energy 
  of about $10^{50}$ erg, ejecta mass around $6 M_{\odot}$, and 
  presupernova radius near $85 R_{\odot}$. 
We confirm the previous result by Turatto et al. (\cite{tmy98}) that 
  the ejecta contain a very low amount of radioactive $^{56}$Ni 
  ($\approx 0.002 M_{\odot}$). 
Modelling the nebular spectrum supports the hydrodynamical model and 
  permits us to estimate the mass of freshly synthesized oxygen (0.02--0.07 
  $M_{\odot}$). 
Combined with the basic results of stellar evolution theory the obtained 
  parameters of SN~1997D imply that the progenitor 
  was a star from the 8--12 $M_{\odot}$ mass range at the main sequence. 
The fact that at least some progenitors from this mass range give rise to 
  core-collapse supernovae with a low kinetic energy 
  ($\approx 10^{50}$ erg) and low amount of radioactive $^{56}$Ni 
  ($\approx 0.002 M_{\odot}$) has no precedent and imposes important 
  constraints on the explosion mechanism. 
We speculate that the galactic supernovae 1054 and 1181 could be attributed to 
  SN~1997D-like events.
\keywords{supernovae: general  --  
          supernovae: individual: SN 1997D -- 
          interstellar medium: supernova remnants}
\end{abstract}
\section{Introduction}
The Type II supernova (SN) 1997D discovered on Jan. 14.15 UT 
  (De Mello \& Benetti 
  \cite{db97}) is a unique event characterized by extremely low expansion 
  velocity, low luminosity, and very low amount ($\approx 0.002 M_{\odot}$) of
   radioactive $^{56}$Ni (Turatto et al. \cite{tmy98}). 
An analysis of the observational data led Turatto et al. (\cite{tmy98}) 
  to conclude that they caught SN~1997D around day 50 after it had exploded 
  as a red supergiant with the mass of 26 $M_{\odot}$ and radius of 
  $R_0\approx 300 R_{\odot}$.
The derived ejecta mass is $M\approx 24 M_{\odot}$ and kinetic energy 
  is $E\approx 4 \times 10^{50}$ erg.
They propose a scenario in which the low $^{56}$Ni mass in SN~1997D is caused 
 by a fall-back of
  material onto the collapsed remnant of the explosion of a 25--40 $M_{\odot}$
  star. 
An exciting implication is that SN~1997D might be accompanied by black hole 
  formation (Zampieri et al. \cite{zsc98}).

Here we present arguments for an alternative view on the origin of 
  SN~1997D, which in our opinion was a descendant from the low end of
  the mass range of core-collapse supernova (CCSN) progenitors.
The problem we attempted to solve first was to find the
  hydrodynamical model, which could reproduce the light curve,
  the velocity at the photosphere, and the line profiles of major strong lines 
  (Section~\ref{sec-photo}).
We emphasize the importance of the line profile analysis, since it provides a 
  robust information on the velocity at the photosphere. 
The latter is a crucial parameter for constraining hydrodynamical models.
Unexpectedly for us it turned out that Rayleigh scattering in SN~1997D is 
  significant and may be used as a powerful diagnostic tool. 
The emphatic role of Rayleigh scattering in this case
 is related to the low energy-to-mass ratio ($E/M$) of SN~1997D (Turatto 
  et al. \cite{tmy98}), which results in the higher than normal 
  density at the photospheric epoch for SN~II-P.
A combination of hydrodynamical modelling and robust analysis of spectra at 
  the photospheric epoch permitted us to impose tight constraints on 
  $E$, $M$, and $R_0$ of SN~1997D.

In addition, we analyzed nebular spectra of SN~1997D using a nebular model 
   (Section ~\ref{sec-nebular}).
To make such an analysis as secure as possible we first checked the model 
  taking advantage of the well studied SN~1987A at a similar epoch.
We found modelling the nebular spectrum of SN~1997D beneficial in 
   discriminating between low and high-mass options for the ejecta.
To our knowledge the present paper is a first attempt of a SN~II-P 
  study to simultaneously make use of all data: light curve, 
  photospheric and nebular spectra.
Some implications of the low mass and kinetic energy of the 
   SN~1997D ejecta for 
  the systematics of CCSN, explosion mechanism, and galactic population of 
  supernova remnants (SNR) are discussed in the final section.
Below we adopt for SN~1997D the dust extinction $A_B=0.0$ mag
  and the distance 13.43 Mpc following Turatto et al. (\cite{tmy98}).

\begin{figure}
  \resizebox{\hsize}{!}{\includegraphics{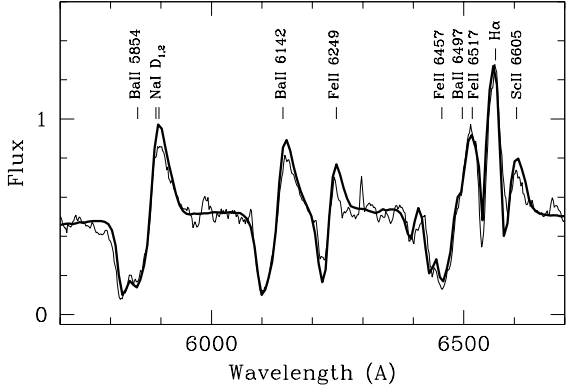}}
  \caption[]{
  Synthetic (thick line) and observed (thin line) spectrum of SN~1997D of 
  Jan. 17 1997.
  All the strong lines are indicated. Note that the model overproduces
  emission in Na I D$_{1,2}$ and Ba II 6142 \AA\ lines.
  }
  \label{fig1} 
\end{figure}

\section{Photospheric epoch}
  \label{sec-photo}
\subsection{The velocity at photosphere}
  \label{sec-photo:vel}

According to general results of hydrodynamical simulations of SNe~II-P
 the radiation cooling of the expanding envelope at the 
 plateau phase proceeds in a specific regime of the cooling 
 recombination wave (Grassberg et al. \cite{gin71}).
As a result the photosphere in SN~II-P resides at the well defined 
 jump between the almost completely
  recombined (ionization degree $\leq 10^{-4}$) transparent atmosphere and 
  the fully ionized sub-photospheric layers of high opacity. 
The velocity at the photosphere determined from the observed scattering 
  line profiles during photospheric epoch (plateau) thus gives us a 
  position of the cooling
 recombination wave and therefore is of vital importance for 
  constraining parameters of the hydrodynamical model. 

To measure the photospheric velocity in the Jan. 17 spectrum of SN~1997D we 
  concentrated on the 5600--6700 \AA\ band which contains strong, clearly-cut 
  spectral lines (Fig.~\ref{fig1}) of
  H, Na~I, Ba~II, Fe~II, and Sc~II. 
Most of them are well observed in other SNe~II-P. 
However, due to the low expansion velocity it is possible to distinguish 
  here some spectral features never observed before, e.g. Sc~II 6605 \AA\ line
  (Fig.~\ref{fig1}).  
A Monte Carlo technique used for modelling the spectrum 
  (Fig.~\ref{fig1}) suggests an absorbing photosphere and a line scattering
  atmosphere (Schwarzschild-Schuster model). 
In total 19 lines are included for this spectral range.
The Sobolev optical depth was computed assuming the analytical
  density distribution in the envelope

\begin{equation}
  \rho = \rho_0\left[1+(v/v_{\rm k})^n\right]^{-1}
  \label{eq:rho}
\end{equation}

\noindent 
  which corresponds to a plateau at velocities $v < v_{\rm k}$ and 
  a steep slope $\rho \propto v^{-n}$ ($n \approx 8$) in the outer layers 
  at $v > v_{\rm k}$. 
Parameters $\rho_0$ and $v_{\rm k}$ are defined by the ejecta mass $M$, 
  kinetic energy $E$, and index $n$. 
The case shown in Fig.~\ref{fig1} is characterized by $M=6 M_{\odot}$, 
  $E=1.2 \times 10^{50}$ erg, and $n=8$, although one may easily fit the 
  spectrum using higher mass and higher energy. 
Since we do not solve the full problem of radiation transfer in ultraviolet 
  we adopt that metals are singly ionized and find level populations
  assuming appropriate excitation temperature ($4200$ K). 
For the standard abundance we assume here, Sc~II 6605 \AA\ line is
  too strong; therefore its abundance is reduced by a factor of two. 
It may well be that the odd behavior of Sc~II line reflects different 
  excitation conditions for Sc~II and other metals rather than abundance 
  pattern.
We found that hydrogen excitation has to be cut beyond $v=1400$ km s$^{-1}$ 
  to prevent washing out the 6500 \AA\ peak.
Close to the photosphere within the layer $\Delta v=300$ km s$^{-1}$
  the net emission in H$\alpha$ is comparable to its scattering component.
We simulated this emission assuming the line scattering albedo greater 
  than unity. 

In spite of its simplicity the model is appropriate for the confident 
  estimate of the photospheric velocity which was found to be $v_{\rm p}=900$
  km s$^{-1}$ (Fig.~\ref{fig1}) with an uncertainty less than 100 km s$^{-1}$. 
The value of $v_{\rm p}=970$ km s$^{-1}$ reported by 
  Turatto et al. (\cite{tmy98}) is consistent with the above estimate. 
Our choice was a compromise between two possibilities: 
  (1) washing out many observed features in the spectrum 
  if $v_{\rm p} \geq 1000$ km s$^{-1}$, and           
  (2) producing significant excess in emission components for Na~I D$_{1,2}$ 
  and Ba~II 6142 \AA\ lines if $v_{\rm p} \leq 800$ km s$^{-1}$. 
The value $v_{\rm p}=900$ km s$^{-1}$, although optimal, still leads to some
  extra emission in Na~I D$_{1,2}$, Ba~II 6142 \AA\, and Fe~II 6249 \AA\ 
  lines (Fig.~\ref{fig1}).
Preliminary analysis indicated that this drawback of the model  may be 
  overcome, if Rayleigh scattering on neutral hydrogen is taken into account.

\subsection {Rayleigh scattering effects}
  \label{sec-photo:Ray}

Rayleigh scattering on neutral hydrogen in the optical dominates over
  Thomson scattering at an extremely low ionization degree $x < 0.001$ which 
  is the case for SN~II-P atmospheres at the photospheric epoch.
To get an idea of the role of Rayleigh scattering in the spectrum of
  SN~1997D we adopt the analytical density profile given by
  Eq.(\ref{eq:rho}) with a power index $n=8$.
Let us first estimate the Rayleigh optical depth $\tau_{\rm R}$
  using the cross-section by Gavrila 
  (\cite{g67}) and assuming conditions of the atmosphere of a normal SN~II-P
  (e.g. SN~1987A) for two extreme cases: completely 
  mixed and unmixed envelopes.
Assuming for SN~1987A $E=1.1 \times 10^{51}$ erg, $M=15 M_{\odot}$, 
   a helium/metal core mass $M_{\rm c}=4.2 M_{\odot}$, and 
  $v_{\rm p}=2600$ km s$^{-1}$ at the age $t=50$ d 
  (Woosley \cite{w88}; Shygeyama \& Nomoto \cite{sn90}; Utrobin \cite{viu93})
  one gets at the wavelength 6142 \AA\ (Ba II line) a Rayleigh optical depth
  of 0.07 (mixed) and 0.1 (unmixed).

\begin{figure}
  \resizebox{\hsize}{!}{\includegraphics{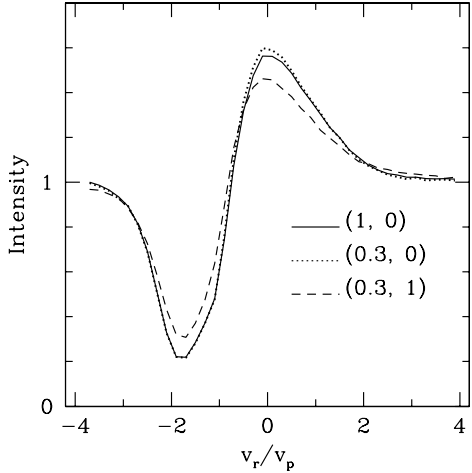}}
  \caption[]{
  Influence of Rayleigh scattering and diffuse reflection from photosphere
  on the resonance line profile. 
  In parentheses are given the thermalization parameter in the 
  sub-photospheric layers and the Rayleigh optical depth in the atmosphere.
  }
  \label{fig2} 
\end{figure}

SN~1997D is essentially different in that respect. 
Adopting the ejecta model by Turatto et al. (\cite{tmy98}), 
  viz. total mass $M= 24 M_{\odot}$, 
  helium/metal core mass $M_{\rm c}=6 M_{\odot}$, $v_{\rm p}= 900$ km s$^{-1}$
  at the expansion time $t=50$ d (the epoch of Jan. 17) one finds
  the Rayleigh optical depth in the range 1.3--1.8 at 
  $\lambda=6142$ \AA, more than one order of magnitude exceeding that in normal
  SNe~II-P. 
For the model with parameters scaled-down by a factor of four 
  ($M=6 M_{\odot}$, 
  helium/metal core mass $M_{\rm c}=1.5 M_{\odot}$), one obtains 
  $0.33 < \tau_{\rm R} < 0.44$. 
Our study showed that such values cannot be ignored in modelling 
 line profiles.

Moreover, to treat Rayleigh scattering in an adequate way one has to 
  abandon the assumption of a fully absorbing photosphere and instead include 
 a diffuse reflection of photons from the photosphere.
We describe the diffuse reflection by a 
  plane albedo  $A(\mu,\epsilon)$ which is a function 
  of cosine $\mu$ of incident angle and thermalization parameter 
  $\epsilon=k_{\rm a}/(k_{\rm a}+k_{\rm s})$. 
Here $k_{\rm a}$ is the 
  absorption coefficient and $k_{\rm s}$ is the scattering coefficient.
In the approximation of the isotropic scattering the plane albedo 
 reads

\begin{equation}
  A(\mu,\epsilon)=1-\phi(\mu,\epsilon) \sqrt{\epsilon}
  \label{eq:A} 
\end{equation}

\noindent 
  where the function $\phi(\mu,\epsilon)$ is defined by the 
   integral equation (cf. Sobolev \cite{s75}) 

\begin{equation}
  \phi(\mu,\epsilon)=1+\frac{1}{2}(1-\epsilon)\mu\phi(\mu,\epsilon)\int_0^1 
       \frac{\phi(\mu_1,\epsilon)}{\mu+\mu_1}d\mu_1 ,
  \label{eq:phi} 
\end{equation}

\noindent 
which was solved numerically to create a table of 
  $A(\mu,\epsilon)$. 

In the absence of Rayleigh scattering, non-zero albedo for $\epsilon=0.3$ 
  slightly (by 4\%) increases the intensity of the emission component 
  compared to the purely absorbing photosphere (Fig.~\ref{fig2}).
The difference obviously becomes larger for smaller thermalization parameter.
Rayleigh scattering significantly decreases the emission component due to 
  backscatter and subsequent absorption of photons by the photosphere in 
  the case of  $\epsilon=0.3$ and $\tau_{\rm R}=1$.
Another effect of Rayleigh scattering is washing out of the absorption trough 
  by continuum photons drifted from blue to red; this effect is especially 
  pronounced for weak lines and is of minor importance for strong lines. 
This modelling shows how the emission excess in  Na~I and Ba~II
  lines (Fig.~\ref{fig1}) may be suppressed.
 
Apart from Rayleigh scattering and diffuse reflection by the photosphere 
  we made two other essential modifications to our Monte Carlo model of line 
  formation. 
First, we took electron scattering into account.
The electron density distribution is recovered from the H$\alpha$ line profile
  using a two-level plus continuum approximation.
Second, we calculated the population of three lowest levels of Ba II using 
  the observed flux in the spectrum on Jan. 17. 
This approximation is fairly good in analyzing the blue side of the absorption 
  trough of the Ba II 6142 \AA\ line. 
We adopted the standard barium abundance (Grevesse \& Sauval \cite{gs98}) and 
  the Ba~II fractional ionization $n({\rm BaII})/n({\rm Ba})=1$. 
The latter seems to be a good approximation for the outer layers of 
  SN~1987A at the stage when strong Ba~II lines are present (Mazzali et al. 
  \cite{mlb92}).

\begin{figure}
  \resizebox{\hsize}{!}{\includegraphics{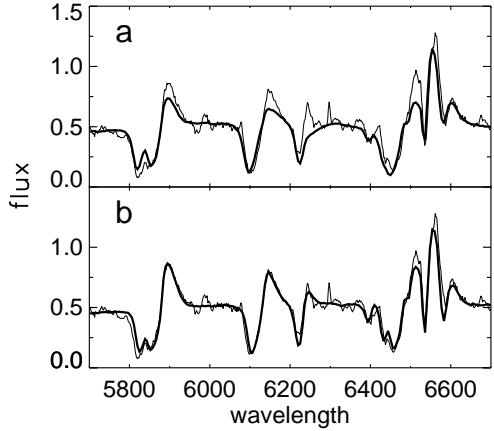}}
  \caption[]{
  Synthetic spectra (thick lines) with  Rayleigh scattering and
  diffuse reflection from the photosphere taken into account, and observed 
  spectrum of SN~1997D on Jan. 17 (thin line).
  The high-mass case (a) shows strong smearing of emission components, 
  while the low-mass case (b) with moderate Rayleigh optical depth 
  fits observations better.
  }
  \label{fig3} 
\end{figure}

With the modified Monte Carlo model the synthetic spectrum is calculated for 
  two relevant cases: a high-mass model with parameters 
  $M=24 M_{\odot}$ and $M_{\rm c}=6 M_{\odot}$ (Fig.~\ref{fig3}a) 
  and a low-mass model with parameters $M=6 M_{\odot}$ and 
  $M_{\rm c}=1.5 M_{\odot}$ (Fig.~\ref{fig3}b). 
Note, that both models have the same photospheric velocity $v_{\rm p}= 900$ 
  km s$^{-1}$ and the same ratio $E_{50}/M=1/6$, where $E_{50}$ is 
  the kinetic energy in units of $10^{50}$ erg and $M$ in $M_{\odot}$.
Complete mixing, which implies a minimum Rayleigh optical depth, gives 
  $\tau_{\rm R}=1.3$ and $0.33$ for high and low-mass models, respectively.  
The thermalization parameter $\epsilon$ in sub-photospheric layers is 0.35 and
  0.24 for high and low-mass models, respectively.
In the high-mass model Rayleigh scattering suppresses the emission components 
  of Na~I D$_{1,2}$, Ba~II 6142 \AA, and Ba~II/Fe~II peak at 6500 \AA\ down 
  to an unacceptably low level (Fig.~\ref{fig3}a).
The low-mass case fits the observations fairly well (Fig.~\ref{fig3}b).
Computations of spectra for different values of Rayleigh optical depth led 
  us to conclude that the tolerated upper limit is 0.6. 
Yet the Rayleigh optical depth cannot be lower than $\approx 0.3$,  
  otherwise the emission in the Na~I D$_{1,2}$
  and Ba~II 6142 \AA\ lines becomes too strong.
We find an optimal value is  $\tau_{\rm R}=0.45$ with an uncertainty of 
  about 0.15 for the Jan. 17 spectrum.

\begin{figure}
  \resizebox{\hsize}{!}{\includegraphics{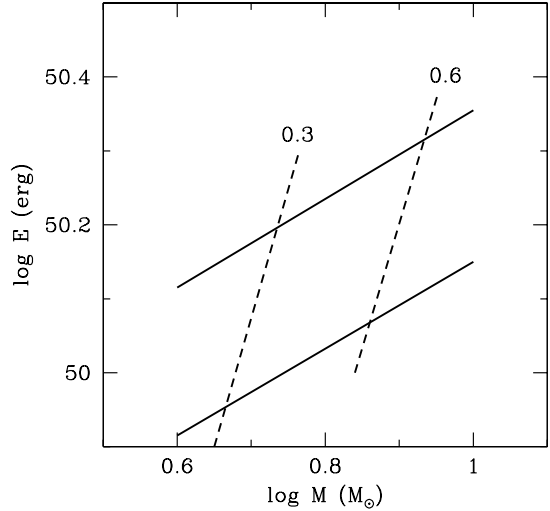}}
  \caption[]{
  The `mass--energy'' diagram for SN~1997D. 
   Solid lines bound the "barium" strip derived from the 
  observed Ba II 6142 \AA\ absorption; dashed lines are
  loci of the constant Rayleigh optical depth with indicated
  values (see Section~\ref{sec-photo:dia}).
  }
  \label{fig4}
\end{figure}

\subsection{Diagnostics of ejecta mass and kinetic energy}
  \label{sec-photo:dia}

The observational limitations upon Rayleigh optical depth in the atmosphere 
  of SN~1997D may be combined with the restriction on the density in the outer
  layers imposed by the blue absorption edge of Ba~II 6142 \AA\
  in order to constrain ejecta mass and kinetic energy. 
The idea may be illustrated using a toy model, in which the supernova envelope 
  is represented by a homogeneous sphere with the boundary velocity $v_0$.
Given the photospheric velocity and Rayleigh scattering optical depth one 
  finds the product $\rho v_0t$, whereas the blue edge of the Ba~II 6142 \AA\
  absorption gives the outer velocity $v_0$. 
For a specific phase $t$ one gets then ejecta mass 
  $M=(4\pi/3)\rho(v_0t)^3$ and kinetic energy $E=(3/10)Mv_0^2$. 

In practice we used a more realistic density profile given by 
  Eq.(\ref{eq:rho}) with a power index $n=8$. 
In this case, likewise for the simple model considered above, one can find, 
  accepting a certain ejecta mass, the corresponding value of the kinetic 
  energy compatible with the blue edge of the Ba~II 6142 \AA\ absorption in 
  the SN~1997D spectrum on Jan. 17. 
Again, we adopted a standard barium abundance with the Ba II ion as the 
  dominant ionization state. 
Variation of the model mass under the condition that the Ba~II 6142 \AA\ 
  absorption is reproduced results in the corresponding variation of the 
  kinetic energy. 
Taking into account uncertainties of the Ba~II 6142 \AA\ absorption
 fit we found a region of allowed parameters
  (``barium'' strip) in the mass--kinetic energy 
 ($M-E$) plane (Fig.~\ref{fig4}).
The lower and upper limits of Rayleigh optical depth, 0.3 and 0.6, 
  respectively, produce another strip of allowed parameters 
  (``Rayleigh'' strip) in this plane. 
The overlap of ``barium'' and ``Rayleigh'' strips gives a tetragonal region 
  where the ejecta mass and kinetic energy of SN~1997D are confined.
One sees that optimal values of ejecta mass should reside around 
   $M\sim 6 M_{\odot}$, 
  while kinetic energy should be close to $E\sim 10^{50}$ erg. 

The suggested diagnostics, unlikely useful for ordinary 
  SNe~II-P,  proved efficient for constraining parameters of SN~1997D.
A warning should be kept in mind that a cosmic barium abundance was assumed 
  here.
This may in general not be the case since SN~1987A demonstrates that barium  
  overabundance in SNe~II-P may be as large as a factor of two relative 
  to the cosmic value  (Mazzali et al. \cite{mlb92}). 
If barium abundance in SN~1997D ejecta is twice the cosmic value, then the 
  ``barium'' strip in the $M-E$ plane has to be shifted down 
  by a factor $\approx 1.3$ towards lower values 
  of kinetic energy.
It is remarkable that this diagnostics does not depend on the supernova
  distance. 
However there is a weak dependence on reddening via the colour temperature 
  determined from 4500 \AA/6140 \AA\ flux ratio which affects the Ba II 
  excitation.
Unaccounted reddening leads to the overestimation of the mass obtained 
  from the Ba II line.

\begin{figure}
  \resizebox{\hsize}{!}{\includegraphics{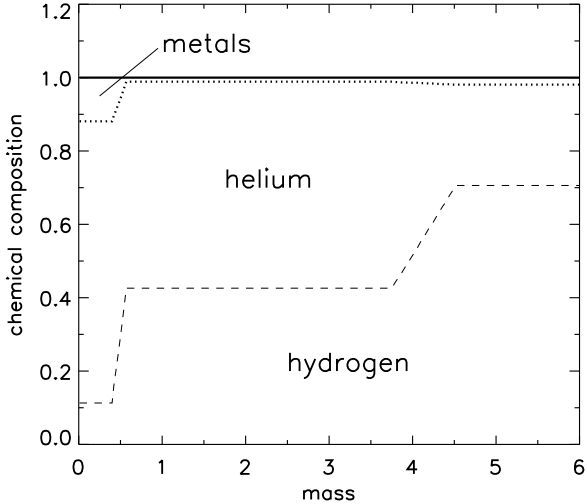}}
  \caption[]{
  Chemical composition of ejecta adopted for the hydrodynamical
  model of SN~1997D. 
  The hydrogen-rich envelope is mixed with the He shell in the inner 4.5 
  $M_{\odot}$ of ejecta. 
  Note that the ejecta mass does not include the collapsed core 
  with a barionic mass of 1.4 $M_{\odot}$.
  }
  \label{fig5} 
\end{figure}

\subsection {Light curve}
  \label{sec-photo:lig} 

The light curve of SN~II-P during the plateau phase is determined by 
  the ejecta mass 
  $M$, kinetic energy $E$, pre-SN radius $R_0$, the structure of the outer 
  layers, the $^{56}$Ni mass and its distribution, and the 
  chemical composition of 
  the envelope (Grassberg et al. \cite{gin71}; Utrobin \cite{viu89}, 
  \cite{viu93}). 
The $^{56}$Ni mass in SN~1997D is reliably measured by the light curve tail.
The structure of outer layers of the pre-SN may normally be recovered 
  from the
  initial phase of the light curve, which was unfortunately missed in the case 
  of SN~1997D. 
Therefore we used a standard pre-SN density structure with the polytrope index 
  of three, though models with other density structure were also tried. 
The abundance of the deeper part of the envelope, e.g., the transition region 
  between the H-rich envelope and metal/helium core affects the final 
  stage of photospheric regime and may therefore be probed by the light curve 
  at the end of the plateau phase. 
In general, the parameters $M$, $E$, and $R_0$ then may be found from 
   the plateau phase duration, luminosity at plateau phase (e.g., in $V$ band), 
   and velocity at the photospheric level. 
In a situation when the plateau phase duration is unknown 
   the optimal Rayleigh optical depth in the atmosphere 
  ($\tau_{\rm R}\approx 0.45\pm0.15$) provides the missing constraint. 
The description of the 
  radiation hydrodynamics code used for supernova study may 
  be found elsewhere (Utrobin \cite{viu93}, \cite{viu96}).

\begin{figure}
  \resizebox{\hsize}{!}{\includegraphics{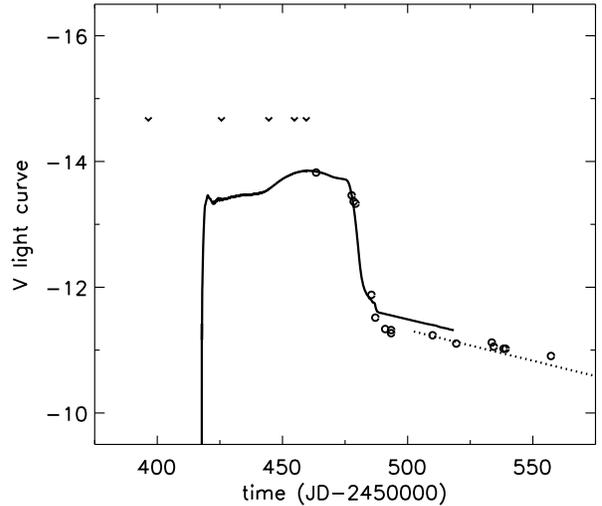}}
  \caption[]{
  Calculated light curve (thick line ), observed photometric data
  (open circles), and upper limits (v-like symbols).
  Data are from Turatto et al.(\cite{tmy98}).
  Dotted line shows the calculated $V$ band luminosity assuming the spectrum 
  energy distribution as in the observed spectrum at $t\approx 150$ d.
  }
  \label{fig6} 
\end{figure}

An extended grid of hydrodynamical models of SN 1997D led us to the 
  conclusion that   requirements imposed by the $V$ light curve,
  velocity at the photosphere, and 
  Rayleigh optical depth are consistent with those estimated above
  from the $M$--$E$ diagram.
The optimal hydrodynamical model is characterized by the following parameters: 
  the ejecta mass $M=6 M_{\odot}$, kinetic energy $E=10^{50}$ erg, and  pre-SN 
  radius $R_0 = 85 R_{\odot}$. 
To prevent the emergence of a luminosity spike at the end of plateau phase 
  and to explain the narrow peak of the H${\alpha}$ emission in the 
  nebular spectrum
  on day $\approx 300$, we suggest mixing between the helium layer and 
  the H-rich envelope (Fig.~\ref{fig5}). 
The adopted helium/metal core mass before mixing is $M_{\rm c}=1.5 M_{\odot}$. 
With 0.002 $M_{\odot}$ of radioactive $^{56}$Ni this choice of parameters
  results in a $V$ light curve, which fits the observational data (Turatto 
  et al. \cite{tmy98}) and is consistent with the observational 
  upper limits by Evans at early epochs (Fig.~\ref{fig6}).
The velocity at the photosphere in this model is 830 km s$^{-1}$
  in agreement with that found from the spectrum synthesis.
Remarkably we obtained the same $E/M$ ratio as Turatto et al. 
  (\cite{tmy98}).
In our model the first spectrum on Jan. 17 corresponds to the epoch of 46 days
  after the explosion in good agreement with the 50 days found by Turatto 
  et al. (\cite{tmy98}).

The diffusion approximation used in the hydrodynamical model breaks down at 
  the transition from the plateau to the radioactive 
  tail about $t\approx 65$ d. 
To reproduce the tail, we translated the bolometric luminosity computed in 
  the hydrodynamical model into the $V$ band luminosity using the two 
  assumptions about the spectrum of escaping radiation at the tail phase.
The first one admits that the spectrum is black-body  with the constant 
  effective temperature calculated at $t=65$ d. 
This gives somewhat higher $V$ luminosity compared to observations at the tail
  stage (Fig.~\ref{fig6}). 
An alternative approach assumes that the spectrum of escaping radiation during
  the tail phase is the same as in the observed nebular spectrum at 
  $t\approx150$ d (Turatto et al. \cite{tmy98}).
The latter assumption is more realistic and provides a good fit to
  observations (Fig.~\ref{fig6}).  
This agreement justifies the adopted $^{56}$Ni mass of 0.002 $M_{\odot}$ 
  originally  obtained by Turatto et al. (\cite{tmy98}).

\begin{figure}
  \resizebox{\hsize}{!}{\includegraphics{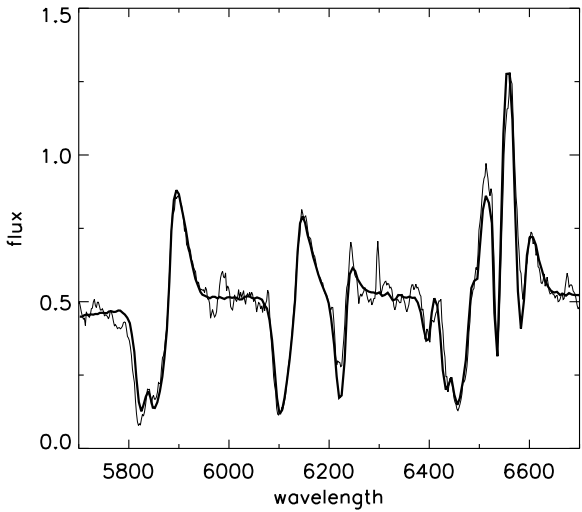}}
  \caption[]{
  Synthetic spectrum for the hydrodynamical model (thick line)
  and observed spectrum on Jan. 17 (thin line).
  }
  \label{fig7} 
\end{figure}

The envelope structure computed in the hydrodynamical model was then used 
  to recalculate the synthetic spectrum in a way similar to that described in 
  Section~\ref{sec-photo:Ray}. 
The model spectrum agrees well with the observed spectrum on Jan. 17 
  (Fig.~\ref{fig7}). 
Of particular importance is an excellent fit for the emission component of 
  the Na~I 5889, 5896 \AA\ doublet which is free of blending, thus being 
  a reliable probe for Rayleigh optical depth in the atmosphere.

Analyzing hydrodynamical models with different sets of input parameters 
  gives us confidence that the envelope mass and the pre-SN radius are 
  determined with the uncertainty of about 1 $M_{\odot}$ and $10~R_{\odot}$, 
  respectively.
Therefore, we estimate the ejected mass as $6\pm1 M_{\odot}$ with the 
  invariant ratio $E_{50}/M=1/6$ and the radius of pre-SN as 
  $85\pm 10~R_{\odot}$.

A dust extinction in the host galaxy (NGC 1536) cannot be ruled out.
It is unlikely, however, significant since the galaxy is nearly face-on.  
With some dust extinction (if any) the kinetic energy and/or the pre-SN 
  radius should be increased accordingly.
For instance, the dust extinction $A_V=0.1$ mag suggests the $13\%$
  increase of the kinetic energy.

\section{Nebular phase}
  \label{sec-nebular}
\subsection{Nebular model}
  \label{sec-nebular:mod}

The high quality late time spectrum of SN~1997D at the nebular epoch 
  $t\approx 300$ d (Turatto et al. \cite{tmy98}) gives us an opportunity
  of a complementary test for the ejecta model. 
Our goal here is to reproduce all the strong lines observed in the spectrum, 
  viz. H${\alpha}$, [O I] 6300, 6364 \AA, and [Ca II] 7291, 7324 \AA\ 
  using the density distribution of the hydrodynamical model.
We assume that the ejected envelope consists of two distinctive regions: 
  a core and an external H-rich envelope. 
The core in the nebular model is a macroscopic mixture of radioactive 
  $^{56}$Ni, H-rich matter (component A), He-rich matter (component B), 
  O-rich matter (component C), and the rest of metals, e.g., C, Ne, Mg 
  (component D). 
The latter does not contribute noticeably to the lines we address.
The density of the A, B, and D components is equal to the model local density 
  $\rho$, while the O-rich matter of density $\rho_{\rm O}$ may be clumpy 
  with the density contrast $\chi_{\rm O}=\rho_{\rm O}/\rho$. 
Voids arising from the oxygen clumpiness are presumably filled in by the 
  $^{56}$Ni bubble material.

The average gamma-ray intensity was calculated using a formal solution of 
  the transfer equation with known distribution of $^{56}$Ni and assuming 
  the absorption approximation with the absorption coefficient $k=0.03$ 
  cm$^2$ g$^{-1}$.
The fraction of deposited energy lost by fast electrons on heating and 
  ionization of hydrogen, helium, and oxygen was taken from Kozma \& Fransson 
  (\cite{kf92}).
The rate of nonthermal excitation and ionization of helium in 
  the H-rich matter 
  was added to the hydrogen ionization rate to take account of hydrogen  
  ionization by UV radiation produced by helium nonthermal excitation and 
  ionization. 
Due to this process the H${\alpha}$\ intensity is insensitive to 
 the He/H ratio.
The photoionization of hydrogen from the second level by 
 hydrogen two-photon radiation and by the   
  Balmer continuum were taken into account as well. 
The Balmer continuum radiation consists of the recombination 
  hydrogen continuum and the rest of ultraviolet radiation 
  created by the radiation 
  cascade of the deposited energy of radioactive $^{56}$Co. 
This additional component of Balmer continuum radiation 
  was specified assuming that a fraction $p$ of the 
  deposited energy is emitted in Balmer continuum with the 
   spectrum $j_{\nu}\propto \nu^{-2}$. 
We adopted $p=0.2$ 
  according to estimates by Xu et al. (\cite{xu92}) for
  SN~1997A at the nebular epoch.

\begin{table}
  \caption{Parameters of nebular models}
  \bigskip
  \begin{tabular}{lcccccc}
  \hline
  Model  & $M$   & $E_{50}$ &  $Z/Z_{\odot}$ & 
  $v_{\rm c}$ (km s$^{-1}$) & $\chi_{\rm O}$\\
  \hline
  TM    &  15  & 11 &  0.4   & 2000  & 5.5 \\
  M1    &  6   & 1 &  0.3   &  600  & 1.2\\
  M2    &  6   & 1 &  1     &  600  &  1\\
  M3    &  24  & 4 &  0.3   &  780  &  1\\
  \hline
  Model &  $M_{\rm c}$ & $M_{\rm H}$ & $M_{\rm He}$  & $M_{\rm O}$ &
  $M_{\rm met}$ \\
  \hline
  TM     & 4.4 & 2    & 0.7  &  1.2   & 0.5   \\
  M1     & 0.64 & 0.12 & 0.45 &  0.035 & 0.035 \\
  M2     & 0.64 & 0.12 & 0.45 &  0.035 & 0.035 \\
  M3     & 4.83 & 0.63 & 0    &  3.4   & 0.8   \\
  \hline
  \end{tabular}
  \label{tab} 
\end{table}

In thermal balance only the principal coolants are included: hydrogen lines, 
  C I 2967 \AA, 4621 \AA, 8727 \AA, 9849 \AA, Mg II 2800 \AA,  
  [O I] 6300, 6364 \AA, [Ca II] 3945 \AA, 7300 \AA, 
  Ca II 8600 \AA, and Fe II lines.
For the sake of simplicity the total Fe II cooling rate of permitted and 
  semi-forbidden lines is assumed to be equal to the cooling rate of one 
  Mg II 2800 \AA\ line. 
However, unlike for the real Mg II line collisional saturation is omitted to
  allow photon branching in Fe II lines.   
Cooling via the excitation of Fe~II forbidden lines is represented by 
  [Fe~II] 8617 \AA\ and [Fe~II] 4287 \AA\  lines, which are the most efficient 
  coolants for the relevant temperature and electron density. 
We include also adiabatic cooling; it is important in the outer region 
  of the hydrogen envelope. 
Metals with a low ionization potential (Mg, Ca, Fe) are assumed singly 
  ionized.

With a specified density distribution and $^{56}$Ni mass the primary fitting 
  parameter is the velocity at the core boundary $v_{\rm c}$, which affects  
  line intensities via the mass of the mixed core $M_{\rm c}$ exposed to the 
  intense gamma-rays.
The amount of matter in components A, B, and C should then be determined  
  from the spectrum fit. 

\begin{figure}
  \resizebox{\hsize}{!}{\includegraphics{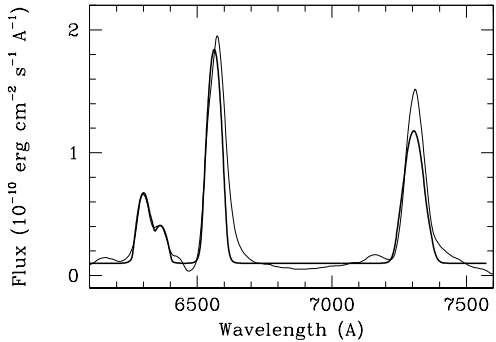}}
  \caption[]{
  Computed (thick line) and observed (thin line) emission lines in 
  the SN~1987A spectrum on day 339 (Pun et al. \cite{pun95}).
  }
  \label{fig8} 
\end{figure}

\subsection{Model test}
  \label{sec-nebular:test} 

Before applying the nebular model to SN~1997D it is instructive to 
  compute the nebular spectrum of the well studied SN~1987A.
We used the CTIO spectrum corrected for reddening at the epoch 339 days
  (Phillips et al. \cite{phh90}; Pun et al. \cite{pun95}).
The primordial metal abundance (Z) is assumed to be 0.4 solar. 
The density distribution in the envelope is approximated by Eq.~(\ref{eq:rho}) 
  with  $n=8.5$. 
Compromise values of ejecta mass $M=15 M_{\odot}$ and kinetic energy
  $E=1.1 \times 10^{51}$ erg are adopted (Woosley \cite{w88}; 
  Shygeyama \& Nomoto \cite{sn90}; Utrobin \cite{viu93}). 
Apart from the $^{56}$Ni mass (0.075 $M_{\odot}$), we specify the amount of  
  metals in the core $M_{\rm met}=0.5 M_{\odot}$ (component D),
  in line with the expectations for an 18--22 $M_{\odot}$ progenitor 
  (Woosley \& Weaver \cite{ww95}; Thielemann et al. \cite{tnh96}). 

A satisfactory description of line profiles and intensities of H${\alpha}$, 
  [O I] 6300, 6364 \AA, and [Ca II] 7291, 7324 \AA\ (Fig.~\ref{fig8}) 
  is obtained with the test model (TM) for a sound choice of parameters 
  (Table~\ref{tab}). 
The table gives ejecta mass ($M$), kinetic energy ($E_{50}$),
  the primordial-to-solar metal abundance ratio 
  ($Z/Z_{\odot}$), velocity at the outer boundary of the mixed 
  core ($v_{\rm c}$), oxygen density contrast ($\chi_{\rm O}$),
  core mass ($M_{\rm c}$), and other core components, viz. H-rich matter
  ($M_{\rm H}$),  He-rich matter ($M_{\rm He}$), O-rich matter ($M_{\rm O}$),
  and metals ($M_{\rm met}$).
All masses in Table~\ref{tab} are given in solar masses. 
The amount of H-rich and He-rich matter in the mixed core 
  inside 2000 km s$^{-1}$  ($2 M_{\odot}$ and $0.7 M_{\odot}$, 
  respectively) are in good agreement with values advocated 
  by  Kozma \& Fransson (\cite{kf98}).
The rest of newly synthesized helium 
  ($\approx 1 M_{\odot}$) is presumably mixed with the H-rich component.
The oxygen mass and density contrast were found from best 
  ``eye-fit'' of flux of the [O I] doublet.  
The value $\chi_{\rm O}=5.5$ corresponds to the oxygen filling factor 
  $\approx 0.045$, a value earlier found by Andronova (\cite{a92}).

\begin{figure}
  \resizebox{\hsize}{!}{\includegraphics{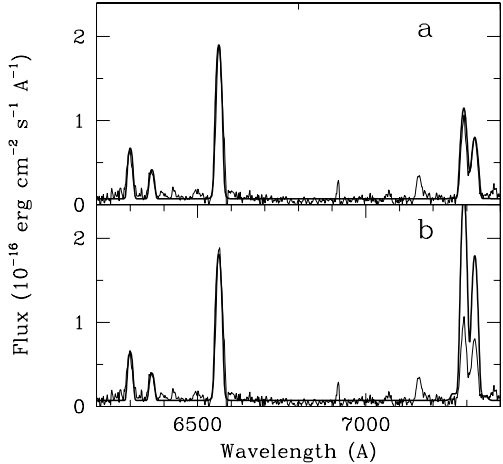}}
  \caption[]{
  Computed (thick line) and observed (thin line) emission lines in SN~1997D 
  spectrum on day 296. 
  Model M1 (a) with primordial metal abundance 0.3 of cosmic describes 
  [Ca II] doublet better than the model M2 (b) with the cosmic abundance.
  }
  \label{fig9} 
\end{figure}

The oxygen mass estimate is hampered somewhat by the uncertainty
  arising from the poorly known fraction of oxygen cooled via CO and 
  SiO emission.
In SN~1987A the mass of cool oxygen in the CO dominated region is estimated 
  as 0.2 $M_{\odot}$ (Liu \& Dalgarno \cite{ld95}). 
With a comparable oxygen mass hidden in the SiO region we thus miss 
  about 0.4~$M_{\odot}$ of oxygen.
Therefore the total oxygen mass must be $\approx 1.6~M_{\odot}$ in rough 
  agreement both with the estimate from the HST spectrum at nonthermal 
  excitation phase (Chugai et al. \cite{nch97}) and predictions of stellar 
  evolution models for an 18--22 $M_{\odot}$ progenitor 
  (Woosley \& Weaver \cite{ww95}; Thielemann et al. \cite{tnh96}). 

Omitting details, we conclude that the test of nebular model in the case of 
  SN~1987A is successful and demonstrates that the model is able to recover 
  reliable values of important parameters.

\subsection{SN~1997D: low and high-mass models}
  \label{sec-nebular:low_high} 

Now we turn to the nebular spectrum of SN~1997D at $t\approx 300$ d.
First, the 6 $M_{\odot}$ case based on the hydrodynamical model
  (Section~\ref{sec-photo:lig}) will be considered. 
Some refinement of the hydrodynamical model is needed, however, to apply it 
  to nebular epoch.
The amount of metals in the mixed core is specified assuming that masses of
  metals and oxygen are equal. 
This is a reasonable assumption for a low-mass pre-SN. 
The oxygen abundance in He-rich matter (component B) was assumed one tenth  
  cosmic, while a carbon abundance of 0.03 is adopted for He-rich matter in the 
  6 $M_{\odot}$ model.
We then also consider the 24 $M_{\odot}$ case based on the model by Turatto 
  et al. (\cite{tmy98}) with the composition taken from Nomoto \& Hashimoto 
  (\cite{nh88}).

The low-mass nebular model of SN~1997D fits the observed spectrum
  fairly well 
  (Fig.~\ref{fig9}a) with the optimal choice of parameters represented by 
  model M1 (Table~\ref{tab}).
The  [Ca II] 7300 \AA\ profile was reproduced for the core velocity 
  $v_{\rm c}=600\pm30$ km s$^{-1}$. 
This parameter is of primary importance, since it determines the 
  absolute mass of the core components with the adopted density structure.
We failed to fit the absolute flux of this line with a cosmic primordial 
  abundance adopted for the model M2 (Table~\ref{tab} and
  Fig.~\ref{fig9}b), while the primordial
  abundance 0.3 of cosmic in model M1 provides an excellent fit.
The oxygen doublet intensity is determined primarily by the mass of the O-rich
  matter, although some 20\% come from He and H-rich matter. 
The value of 0.035 $M_{\odot}$ is corrected for the unseen cool 
  oxygen assuming 
  that we see 3/4 of all the pure oxygen in the [O I] doublet as in SN~1987A. 
A strong oxygen overdensity is not required.  
We found that $\chi_{\rm O}=1.2$ in model M1 provides 
  somewhat better agreement with the observed ratio of [O I] doublet 
  components than $\chi_{\rm O}=1$.
The amount of He-rich matter is a lower limit for the mass of the He shell
  in the pre-SN. 
One may admit up to 1 $M_{\odot}$ of helium mixed microscopically with 
  the H-rich envelope without a notable effect on line intensities.

Unfortunately our nebular model is not applicable to the earlier
   nebular spectrum of SN~1997D on day 150. 
The reason is the 
   significant optical depth in the Paschen continuum predicted by the
   model. 
In such a situation the multilevel statistical equilibrium 
   must be solved together with a full radiation transfer, which is 
   beyond the scope of this paper. 
Moreover, we found that the observed H${\alpha}$ profile at this epoch
  is odd exhibiting a significant redshift of unclear origin. 
A prima face explanation assuming $^{56}$Ni asymmetry cannot accommodate to 
  the late time nebular spectrum ($\approx 300$ d) lacking 
   such an asymmetry.

\begin{figure}
  \resizebox{\hsize}{!}{\includegraphics{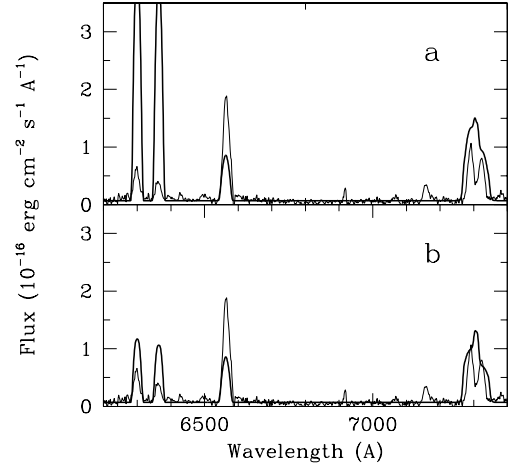}}
  \caption[]{
  Computed nebular spectrum (thick line) in the high-mass case along with the 
  observed (thin line) emission lines in SN~1997D spectrum on day 296. 
  (a) -- model M3 without reduction of oxygen line-emitting mass; 
  (b) -- the same model but with five-fold reduction of the oxygen 
  line-emitting mass.
  Neither of the model spectra fits the data.
  }
  \label{fig10} 
\end{figure}

To evaluate the uncertainty related with the assumption of the same 
  fraction (1/4) 
  of cool O-rich gas as in SN~1987A, we compared parameters relevant to 
  molecular formation (density and temperature) in the O-rich matter at 
  a similar nebular epoch. 
We found that the density of O-rich gas in SN~1997D is lower, while the 
  temperature is somewhat higher compared to SN~1987A. 
Both parameters suggest therefore that formation of molecules in SN~1997D 
  cannot be more efficient than in SN~1987A, which means that the fraction 
  of unseen pure oxygen in SN~1997D does not exceed that in SN~1987A.
With the uncertainty of the core velocity the estimated 
  range of pure oxygen mass in SN~1997D is 0.02--0.07 $M_{\odot}$. 

We applied the nebular model to the high-mass case ($M=24 M_{\odot}$).  
Due to the large mass of the He/O core the velocity of the core boundary is 
  too high and inconsistent with the observed [Ca II] doublet profile. 
Mixing all the freshly synthesized helium with the hydrogen envelope reduces 
  the core velocity but not sufficiently to resolve this controversy 
  (Fig.~\ref{fig10}a).
Another serious problem is a high [O I] doublet flux and wrong doublet ratio. 
The five-fold reduction of amount of line-emitting oxygen, presumably due to 
  molecular formation and cooling, alleviates the problem of total flux in 
  the [O I] doublet. 
Yet the problem of high $I(6364)/I(6300)$ ratio in this model remains 
  (Fig.~\ref{fig10}b).

Summing up, we found that the hydrodynamical model of moderate mass ejecta 
  ($6~M_{\odot}$) which contains low amount of freshly synthesized oxygen 
  (0.02--0.07 $M_{\odot}$) is consistent with the nebular spectra of SN~1997D. 
The model of high-mass ejecta as such is incompatible with the observed 
  nebular spectra.

\begin{figure}
  \resizebox{\hsize}{!}{\includegraphics{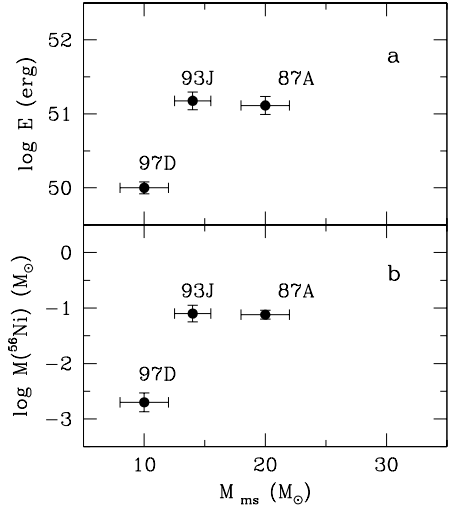}}
  \caption[]{
  Kinetic energy (a) and $^{56}$Ni mass (b) 
 as a function of progenitor mass for 
  three core-collapse supernovae. 
  }
  \label{fig11} 
\end{figure}

\section{Discussion}
  \label{sec-dics} 

We attributed SN~1997D to the SN~II-P event characterized by the 
  kinetic energy $E \approx 10^{50}$ erg and ejecta mass $6 \pm 1
  M_{\odot}$.
The ejecta are dominated by H-rich matter and contains 0.02--0.07 
  $M_{\odot}$ of freshly synthesized oxygen.
The estimated $^{56}$Ni mass is about 0.002 $M_{\odot}$ in accordance with 
  the value found by Turatto et al. (\cite{tmy98}).
The pre-SN had a moderate radius of 85 $R_{\odot}$ and possibly a low 
  primordial metallicity, 0.3 cosmic.

At first glance the suggested low metallicity of SN 1997D 
 disagrees with  the assumed cosmic abundance of barium which 
 implies a relative overabundance by a factor of three. The latter
 should not confuse us, however, after SN~1987A in which the relative  
 barium overabundance is about five  (Mazzali et al. \cite{mlb92})
 for the comparable metallicity of both supernovae.
Amazingly, the relatively small pre-SN radius and low primordial 
 metallicity of SN~1997D both are reminiscent of SN~1987A.  
Possibly it reflects some trend for low metallicity progenitors to have 
  smaller pre-SN radii compared to SNe~II-P with cosmic metallicity.

When combined the ejecta mass and the collapsed core mass (presumably 1.4
  $M_{\odot}$), the total pre-SN mass amounts to 6--9 $M_{\odot}$ prior to 
  outburst.
The main-sequence progenitor likely was more massive because of a possible 
  wind mass-loss.
In the context of general results of stellar evolution theory,
  the low mass of freshly synthesized oxygen ($<0.1~M_{\odot}$) is
  compelling evidence that the progenitor of SN~1997D was a main-sequence star
  from the 8--12 $M_{\odot}$ range.
These stars are known to end their life with very low amount 
  ($<0.1~M_{\odot}$)
  of synthesized oxygen (Nomoto \cite{n84}; Woosley \cite{w86}).
Ejecta of CCSN produced by such stars must contain 
 very small amount of $^{56}$Ni, significantly less than 
  normal CCSN (Woosley \cite{w86}), which is also in line with 
  SN~1997D.

The fact that at least some CCSN originating from the 8--12 $M_{\odot}$
  mass stars have low kinetic energy ($E \approx 10^{50}$ erg) and eject 
  small amounts of $^{56}$Ni ($\approx 0.002 M_{\odot}$)  modifies 
  a picture of CCSN with 
  ``standard'' kinetic energy of $\sim 10^{51}$ erg and $^{56}$Ni mass of 
  0.07--0.1 $M_{\odot}$.
The new situation in the systematics of CCSN is visualized by the
 $E-M_{\rm ms}$ and  $^{56}$Ni mass--$M_{\rm ms}$ plots
 (Fig.~\ref{fig11}) which show the position of 
 SN~1997D along with two other well studied CCSN, 
 SN~1987A (Woosley \cite{w88}; Shygeyama \& Nomoto 
  \cite{sn90}; Utrobin \cite{viu93})
 and SN~1993J (Bartunov et al. \cite{bbpt94}; Shigeyama et al. \cite{setal94};
  Woosley et al. \cite{woetal94}; Utrobin \cite{viu96}).
The primary significance of this plot is that both nearly constant  
 kinetic energy ($\sim 10^{51}$ erg) and $^{56}$Ni mass 
 ($\sim 0.08 M_{\odot}$) in the range of progenitor masses
 between $\approx 13 M_{\odot}$ and $\approx 20 M_{\odot}$
 abruptly drop at the low end of massive star range producing CCSN
 (around 10 $M_{\odot}$).
It would be not unreasonable to consider that SN~1997D is a 
 prototype for a new family of CCSN (below referred to as ``weak CCSN'')
  which occupies the same place on the $E$--$M_{\rm ms}$ and  
  $^{56}$Ni mass--$M_{\rm ms}$ plots as SN~1997D.

Unfortunately, there are no clear theoretical predictions in regard 
  to weak CCSN. 
Yet current trends in the core-collapse modelling seem to 
 be generally  consistent with Fig.~\ref{fig11}.
Two explosion  mechanisms are related to producing CCSN: prompt 
 (core rebound) and delayed (neu\-trino-driven mechanism). 
For 8--10 $M_{\odot}$ progenitors
   the prompt mechanism attains its highest efficiency 
   (Hillebrandt et al. \cite{hnw84}) with the kinetic 
  energy of ejecta  $\sim10^{50}$ erg 
   (Baron \&  Cooperstein \cite{bc90}), while 
  the delayed mechanism, on the contrary, 
  has the lowest efficiency in this mass range with similar
  energy $\sim10^{50}$ erg  (Wilson et al. \cite{wmww86}). 
Thus both explosion mechanisms remain viable in the context of
 SN~1997D. 
However, possibly only the neutrino-driven mechanism is able
 to account for the kinetic energy increase with
 the progenitor mass in the 
 range from about 10 $M_{\odot}$ to $\approx 13 M_{\odot}$
 (Wilson et al. \cite{wmww86}; Burrows \cite{b98}).

How frequent are SN~1997D-like phenomena? 
The first thought is that they are extremely rare, since among 
  $\sim 10^{2}$ identified SN~II-P events only one such case has been
  discovered  so far.
However, with the low absolute luminosity ($\approx -14$ mag) and the brief 
  plateau duration (40--50 days) compared to normal SN~II-P characteristics 
  ($\approx -16.5$ mag and 80--100 days, respectively) it would not be 
  surprising, if SN~1997D-like events were as frequent as $\sim 20\%$ of 
  normal SN~II-P rate. 
Such a rate might be maintained by progenitors from the mass 
 interval $\Delta M_{\rm ms}\approx 1 M_{\odot}$ 
 in the vicinity of main-sequence mass $\approx 10 M_{\odot}$.

Progenitors from the 8--12 $M_{\odot}$ mass range were 
  suggested earlier as counterparts for supernovae
 with a dense circumstellar wind, low ejecta 
  mass ($\sim 1 M_{\odot}$), and possibly normal kinetic energy 
 (e.g. SN~1988Z, Chugai \& Danziger \cite{cd94}). 
The present attribution of SN~1997D to the same mass range introduces some 
  dissonance with the former conjecture. 
In reality, this controversy is not serious since 8--12 $M_{\odot}$ 
  progenitors are characterized by very complicate 
  evolutions of their cores (Nomoto \cite{n84}; Woosley \cite{w86})
  and therefore a different outcome for slightly 
 different initial mass is quite conceivable.
Moreover, it may well be that with a normal metallicity 
 presupernova of weak CCSN also vigorously loses mass
  and explodes in a dense wind thus producing a
  luminous supernova (possibly SN~IIn) due to the ejecta 
  wind interaction. 

Another intriguing possibility is 
 that a presupernova of weak CCSN might lose all the hydrogen 
  envelope in a close binary system before the explosion.
In this case weak CCSN will be a mini-version of SN~Ib with
 low explosion energy, low amount of $^{56}$Ni, and, eventually, 
 low luminosity. 
Unfortunately it will not be easy to detect such events.

If weak CCSN are as frequent as $\sim 20\%$ of 
 all CCSN, a good fraction of galactic population of SNR 
 may be related to these supernovae.
We cannot miss the opportunity to speculate that 
 at least two historical supernovae SN~1054 and SN~1181 might be 
 identified with weak CCSN. 
Nomoto (\cite{n84}) already argued that the Crab Nebula was 
 created by CCSN with the progenitor mass around 9 $M_{\odot}$. 
The luminosity of SN~1054, which was normal for SN~II, could be explained 
  then by the interaction of ejecta with a dense pre-SN wind.
This suggestion in fact is a modification of the earlier idea that 
  the initial phase of the light curve of SN~1054 could be powered by 
  the shock wave propagating in the circumstellar ($r\sim 10^{15}$ cm) 
  envelope (Weaver \& Woosley \cite{ww79}).
The second possible counterpart of a galactic,  weak CCSN is SN~1181. 
With the absolute luminosity of $-13.8$ mag at maximum 
  (Green \& Gull \cite{gg82}) and half year period of visibility 
  SN~1181 is the closest analogue of SN~1997D. 
Low radial velocities ($\leq 1100$ km s$^{-1}$) of SN~1181
  filaments claimed by 
  Fesen, Kirshner \& Becker (\cite{fkb88}) seem to strengthen 
  this identification.
If the association of SN~1181 is correct then we expect to find
 very low amount of newly synthesized oxygen and iron-peak elements
 in this supernova remnant.

\begin{acknowledgements}

We thank Massimo Turatto for sending spectra of SN~1997D.
We are grateful to Ken Nomoto, Bruno Leibundgut, Peter Lundqvist, and
  Wolfgang Hillebrandt for discussions and comments.
N.Ch. thanks Bruno Leibundgut for hospitality at ESO and
  V.U. thanks Wolfgang Hillebrandt and Ewald M\"{u}ller for hospitality
  at MPA.
This work was supported in part by the RFBR (project 98-02-16404) and
  the INTAS-RFBR (project 95-0832).

\end{acknowledgements}
\end{document}